\documentclass[aps,pra,reprint,superscriptaddress,floatfix]{revtex4-1} 
\usepackage{amsmath}
\usepackage{amssymb}
\usepackage{graphicx}

\usepackage{placeins} 
\usepackage{color}
\usepackage[usenames,dvipsnames,svgnames,table]{xcolor}
\usepackage[colorlinks=true,
            linkcolor=blue,
            urlcolor=blue,
            citecolor=blue]{hyperref}
\usepackage[latin1]{inputenc} 
\usepackage{url} 
  
\begin{document}
\title{Selective measurement of charge dynamics in an ensemble of nitrogen-vacancy centers in nano- and bulk diamonds}

\author{R. Giri}
\email[rakshyakar.giri@iit.it]{}
\author{C. Dorigoni}
\author{S. Tambalo}
\author{F. Gorrini}%
\affiliation{Center for Neuroscience and Cognitive Systems, Istituto Italiano di Tecnologia, Corso Bettini 31, Rovereto, 38068, Italy}%
\author{A. Bifone}
\affiliation{Center for Neuroscience and Cognitive Systems, Istituto Italiano di Tecnologia, Corso Bettini 31, Rovereto, 38068, Italy}%
\affiliation{Department of Molecular Biotechnology and Health Sciences, University of Turin, Via Verdi 8, Turin, 10124, Italy}%

\date{\today}

\begin{abstract}
Nitrogen-vacancy (NV) centers in diamond have attracted considerable interest in sensing of weak magnetic fields, such as those created by biological systems. Detecting such feeble signals requires near-surface NV centers, to reduce the distance between NVs and sources. Moreover, dense ensembles of NVs are highly desirable to reduce measurement time. However, robust charge state switching is often observed in these systems, resulting in a complex interplay between charge and spin dynamics that can reduce the attainable level of spin polarization, and consequently, sensitivity. Understanding the mechanisms behind charge state switching is, therefore, crucial to developing NV based sensors. Here, we demonstrate a novel method to selectively measure charge dynamics in an ensemble of NVs by quenching the spin polarization using an off-axis magnetic field. Utilizing this technique, we show that, in nanodiamonds, charge state instability increases with increasing NV density. In the case of bulk single crystal diamond, we show that  NV centers located near the surface are more stable in the neutral (NV$^{0}$) charge state, while the negatively charged (NV$^{-}$) form is more stable in bulk.
\end{abstract}
\maketitle

\section{Introduction}
The Nitrogen-vacancy (NV) center is one of the numerous color centers found in the diamond. It is commonly observed in either the neutral (NV$^{0}$) or negatively charged state (NV$^{-}$) \cite{Davies1976, Davies1992, Iakoubovskii2000}. While the spin state of the negatively charged NV centers can be manipulated with optical and microwave excitation \cite{Gruber1997}, experimental control of the spin state of the NV$^{0}$ centers remains elusive, although theoretically possible \cite{Felton2008, Gali2009}. The electron spin degree of freedom of the NV$^{-}$ centers has been exploited for potential applications in quantum sensing, nanoscale-MRI, and quantum computing \cite{Dolde2011, Childress2013, Rondin2014, Tetienne2016}. Therefore, controlling the charge state of the NV centers is crucial.  A stable and well-controlled NV charge state not only improves the sensitivity of detection, but also leads to applications such as sensing of electrochemical potentials \cite{Karaveli2016}, and enhanced nuclear spin coherence time \cite{Pfender2017}.

  While single isolated NV centers enabled magnetic sensing with nanoscale resolution, ensembles of NV centers are desirable in applications where sensitivity is a critical factor to increase photon counts and reduce measurement time \cite{Rondin2014}. However, charge state instability may become acute in the case of dense ensembles of NVs \cite{Choi2017}.  Various factors, such as nitrogen defects, surface states, vacancies, and other deep level defects can influence the charge state stability of the NV centers during initialization as well as in the dark \cite{Manson2005, Dhomkar2018b, Manson2018}.  In spite of the detrimental impact of charge state instability on spin polarization \cite{XDChen2015a, Choi2017}, and consequently on the sensing capabilities of NV centers, a clear understanding of charge dynamics of NV ensembles is lacking. 
 
Attempts to control the charge states of NVs by manipulating the Fermi level are reported in the literature  \cite{Hauf2011, Grotz2012, Doi2014, Karaveli2016, Maki2018, Murai2018}.  It was demonstrated that an ensemble of NV centers could be stable enough to be used as a charge based data storage medium \cite{Jayakumar2016, Dhomkar2016}. Other studies on dense ensembles of NVs  and near-surface single NVs reported a strong interplay between charge and spin dynamics during illumination as well as in the dark that can interfere with spin measurements \cite{Choi2017,Giri2018, Bluvstein2018}. Disentangling the two contributions is critical to understand the underlying physical mechanisms, and to enable effective control of charge dynamics in many applications.

Selective measurements of the spin dynamics in the dark were shown to be feasible by removing contributions from recombination as well as ionization of NVs in the dark \cite{Jarmola2012, Myers2017, Bluvstein2018}. However, a complementary technique to selectively measure charge dynamics is lacking. To this end, pulse sequences involving green and yellow lasers have been proposed  \cite{Choi2017, Dhomkar2018b, Hopper2018}, but they are not effective to remove the spin contribution entirely. 

Here, we note that an off-axis magnetic field (few degrees off the NV-axis) can mix the $m_s=0$ and $m_s=\pm 1$ spin states of the NV centers, and lead to a reduction of excited state lifetime, fluorescence intensity, as well as electron spin resonance contrast \cite{Epstein2005, Lai2009}.  An external magnetic field along the [100] crystal axis makes an equal angle of 54.7$^\circ$ to all the four NV orientation, and a field around 500 G $\parallel$ [100] could completely mix the spin states resulting in zero spin polarization \cite{Lai2009, Tetienne2012}. 

\begin{figure*}[!ht]
\centering
  \includegraphics[width=1\textwidth]{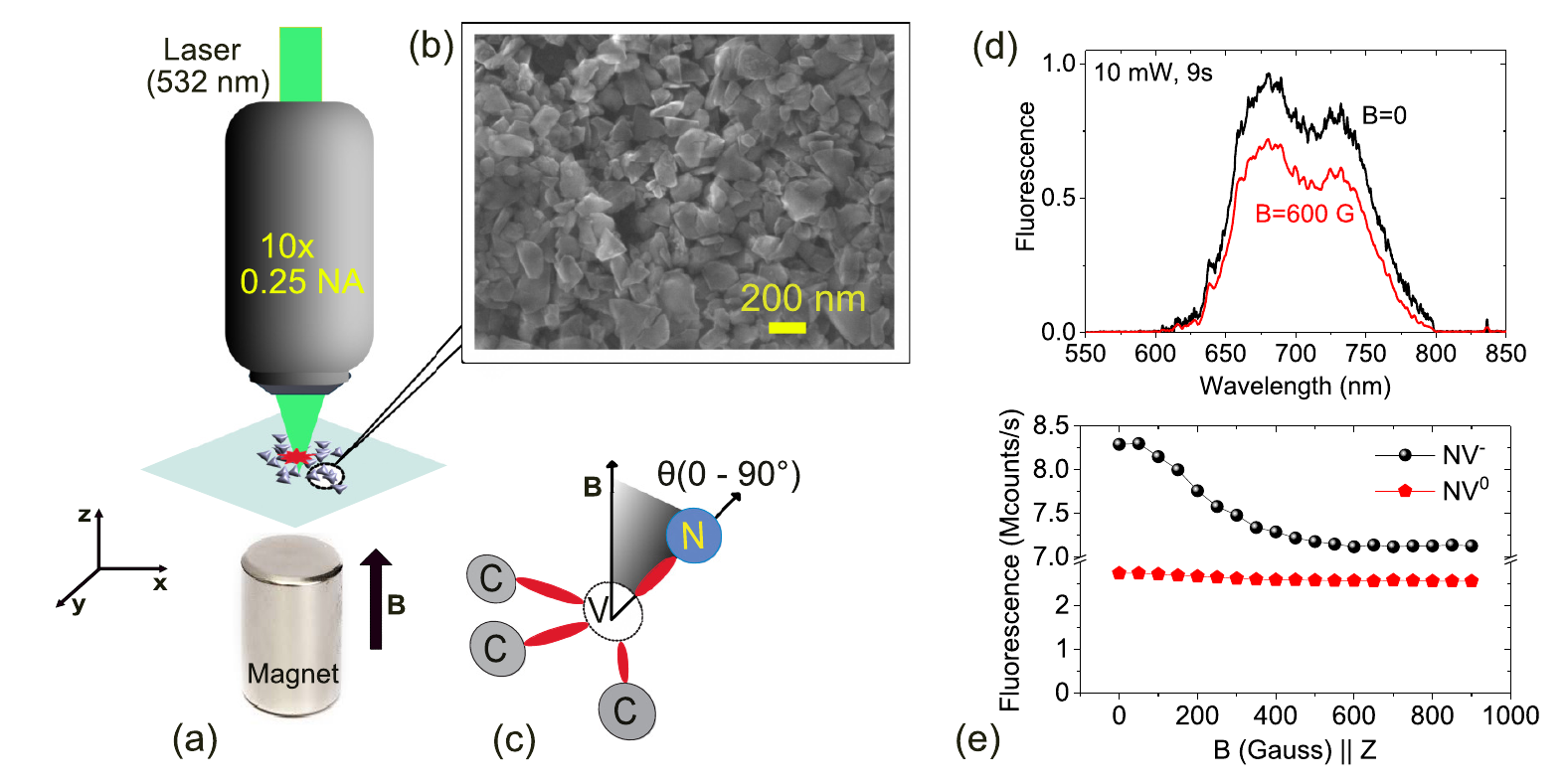}
  \caption{(a) Schematic of the experimental setup. The nanodiamonds (nominally 100 nm diameter) are deposited on a cover glass, and the green excitation laser was focussed on the sample by a 0.25 NA microscope objective. The fluorescence was collected using the same objective, and bandpass filters were used to filter out the fluorescence corresponding to  NV$^{-}$  and NV$^{0}$  emission. A permanent magnet placed directly below the sample produced the static magnetic field. (b) SEM image of the NDs deposited on a silicon substrate showing random crystal orientations. (c) The NV defect axes oriented randomly to the applied magnetic field, and the angle $\theta$ could range from 0 to  90$^\circ$.  (d) Fluorescence emission spectra of the NV ensembles in the nanodiamonds with and without a magnetic field. (e) Variation of NV$^{-}$  and NV$^{0}$  fluorescence intensity as a function of external magnetic field, indicating mixing of spin states, but no apparent change in the ionization of the NVs.}
  \label{figure1a_1d}
\end{figure*} 

In this work, we demonstrate this principle in an ensemble of NV centers in nanodiamond powders and in bulk diamond, showing that it is feasible to selectively measure charge state dynamics during laser illumination as well as in the dark. We observed that increasing the strength of the applied magnetic field results in a reduction of the spin polarization during a 532 nm laser illumination. Complete elimination of the spin polarization was achieved at 600 G, revealing pure ionization-recharge dynamics, without any visible impact of the magnetic field in the NV$^{-}$  and NV$^{0}$ populations. We applied this approach to study charge dynamics in nanodiamonds with different densities of NVs, and in bulk single-crystal diamond as a function of the distance of the NVs from the surface.

\section{Methods}
The nanodiamond (ND) samples used in this study are uncoated fluorescent nanodiamonds from Bikanta (Berkely, CA), with nominal diameter 100 nm (FND100). The nanodiamonds are strongly fluorescent due to a high concentration of NVs ($\approx$ 5 ppm), with each ND containing about 500 NV centers.  We deposited the NDs on a cover glass to carry out all the measurements.

We obtained the chemical vapor deposition (CVD) grown single crystal diamonds used here from Element Six Ltd. One sample (S6) is a single crystal diamond plate,  (100) oriented,  with $<$ 1 ppm nitrogen. We estimated the concentration of NV centers to be about 0.028 ppm by comparison of the level of fluorescence with a reference sample. 
The other single crystal diamond (EG6) is an electronic grade sample, (100) oriented, with nitrogen concentration less than 5 ppb. The sample has been implanted with ${^{15}}$N ions at 15 keV with a dose of 1x10${^{13}}$ ions/cm${^{2}}$, using an angle of incidence of 7$^\circ$. Ion average range, calculated by SRIM \cite{Ziegler2008}, is about 21 nm. Subsequent annealing in helium gas at 850 $^\circ$C for 2 hours results in near-surface NV centers, which we estimated to be about 2.4 ppb.

\begin{figure*}[th]
\centering
  \includegraphics[width=\textwidth]{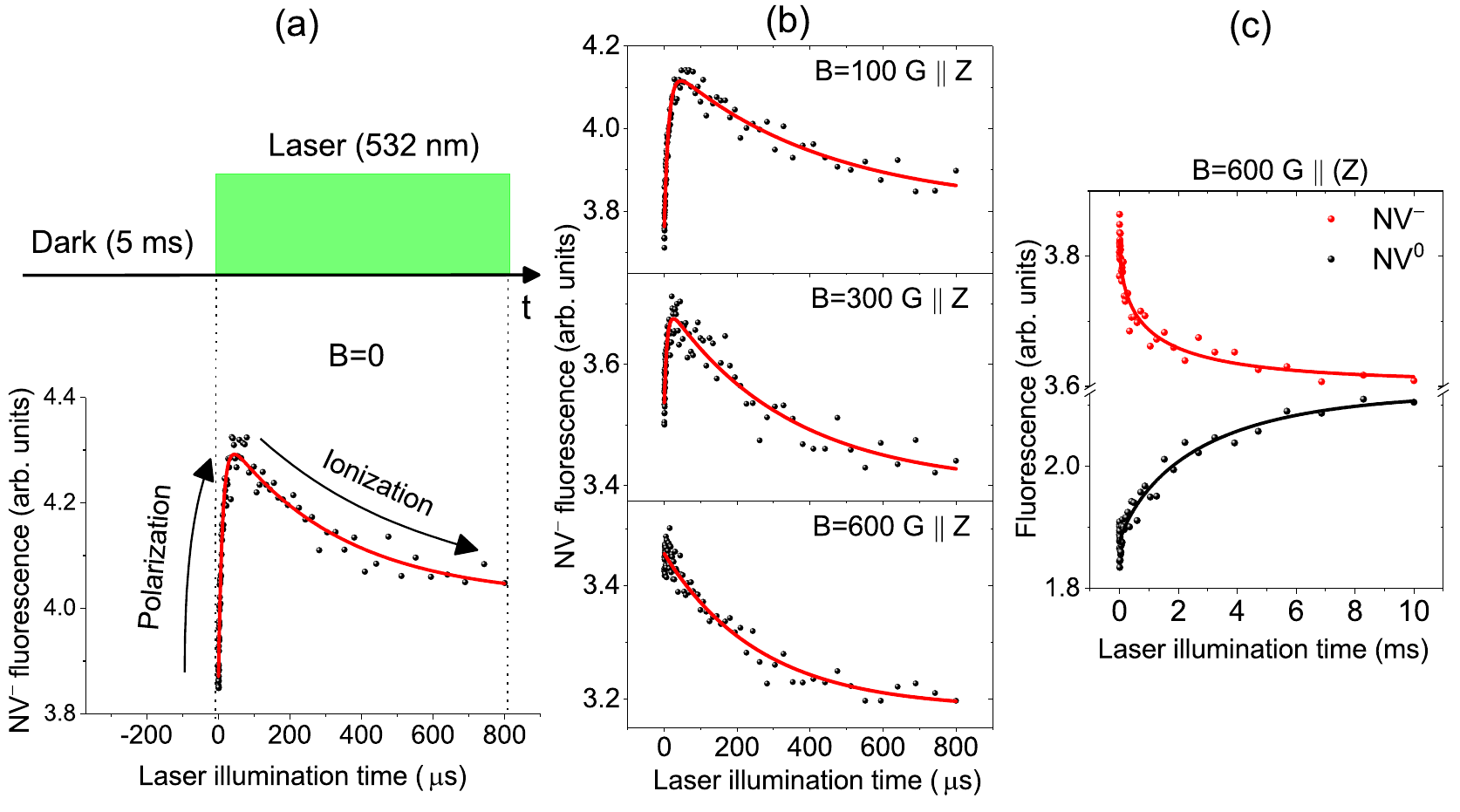}
  \caption{(a) Pulse sequence used to measure the charge and spin dynamics in the nanodiamond powder during 532 nm laser illumination. The system is allowed to relax for 5 ms in the dark, and the time dependence of the fluorescence intensity is recorded during excitation. The lower panel depicts the time evolution of fluorescence intensity at B = 0, indicating a convolution of spin and charge dynamics. (b) Fluorescence signal temporal evolution for increasing applied field strength, showing a gradual reduction of spin polarization. (c) Complete suppression of spin polarization is obtained at B= 600 G $\parallel$ Z, showing pure charge dynamics: ionization of NV$^{-}$ centers is paralleled by an increase of NV$^{0}$ centers. The red and black solid lines represent biexponential fits.}
  \label{figure2a_2c}
\end{figure*} 

We used a home-built confocal microscope (see Fig. \ref{figure1a_1d}a) equipped with an objective lens of 0.25 NA and operating at room temperature. We used a 532 nm laser for both spin and charge state initialization as well as detection. The NV$^{0}$ and NV$^{-}$ centers have zero-phone lines at  575 nm and 637 nm respectively with broad phonon sidebands \cite{Iakoubovskii2000}. We used a series of bandpass filters to detect fluorescence from very narrow regions (550-600) nm and (750-800) nm corresponding to NV$^{0}$ and NV$^{-}$ emission respectively. A single-photon-counting module (Excelitas, SPCM-AQRH-14-FC) was used to detect the fluorescence. An acousto-optic modulator (AA Optoelectronics, MT 200-A0, 5-VIS) produced the excitation laser pulses, and programmable transistor-transistor logic (TTL) pulse generator (Spincore, PulseBlaster ESR-PRO) was used to generate the pulse sequences. A permanent magnet was mounted on a linear translation stage to apply a variable static magnetic field along the vertical (z) axis. 

\section{Results and Discussion}
We show the fluorescence emission spectra of the ensembles of NV centers in the nanodiamond powder sample in Fig. \ref{figure1a_1d}d, obtained by continuously illuminating the sample with a weak 10-mW laser at 532 nm. Only a 532-nm notch filter and 800-nm short pass filters were used for this measurement. The NV$^{0}$ related fluorescence is negligible in this sample.  Application of an external magnetic field, $\bold{B}$  reduced the NV$^{-}$ fluorescence. However, any effect on NV$^{0}$ fluorescence level is not obvious. To get a clear picture of the effect of $\bold{B}$, we spectrally filtered the NV$^{0}$ (550-600 nm band) and NV$^{-}$ (750-800 band) fluorescence. Figure \ref{figure1a_1d}e shows the variation of NV$^{0}$ and NV$^{-}$ fluorescence with increasing $\bold{B}$. The NV$^{-}$ fluorescence decreases monotonically with the increase in the strength of $\bold{B}$ and reaches saturation at fields stronger than 500 G. In the case of nanodiamond powder, the crystal orientations are random, and each NV makes a random angle to $\bold{B}$ (Fig. \ref{figure1a_1d}). The monotonic decrease of the fluorescence with increasing $\bold{B}$ suggests that $\theta_{avg.}$$>$ 20$^\circ$ \cite{Tetienne2012}, and is a general indication of sufficient mixing of the $m_s=0$ and $m_s=\pm 1$ spin states \cite{Manson2006, Lai2009, Tetienne2012}. However, the NV$^{0}$ fluorescence level is not significantly affected by the change in the magnetic field strength. These results suggest that the magnetic field quenches only the spin polarization of the NV centers, but does not significantly affect the NV populations in either of the charge states. These results are consistent with previously reported results on an ensemble of NV centers in high-pressure high-temperature (HPHT) bulk diamond \cite{Capelli2017, Manson2018}. We note here that contrasting results were obtained in single NV centers in high purity CVD diamond \cite{XDChen2013}, indicating fundamentally different ionization and recombination mechanisms in NV ensembles. 

\begin{figure}[th]
\centering
  \includegraphics[width=0.5\textwidth]{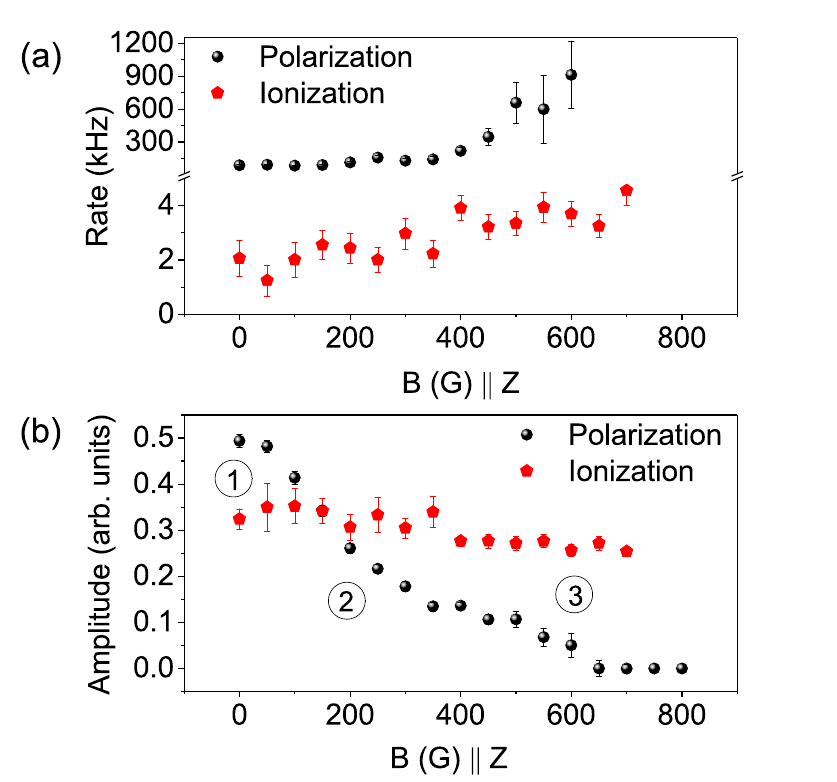}
  \caption{(a) Rates, and (b) amplitudes of spin polarization and ionization of NVs during 532 nm laser illumination, extracted from bi-exponentials fits of curves as in Fig. \ref{figure2a_2c}b.}
  \label{figure3a_3b}
\end{figure} 

\begin{figure}[th]
\centering
  \includegraphics[width=0.5\textwidth]{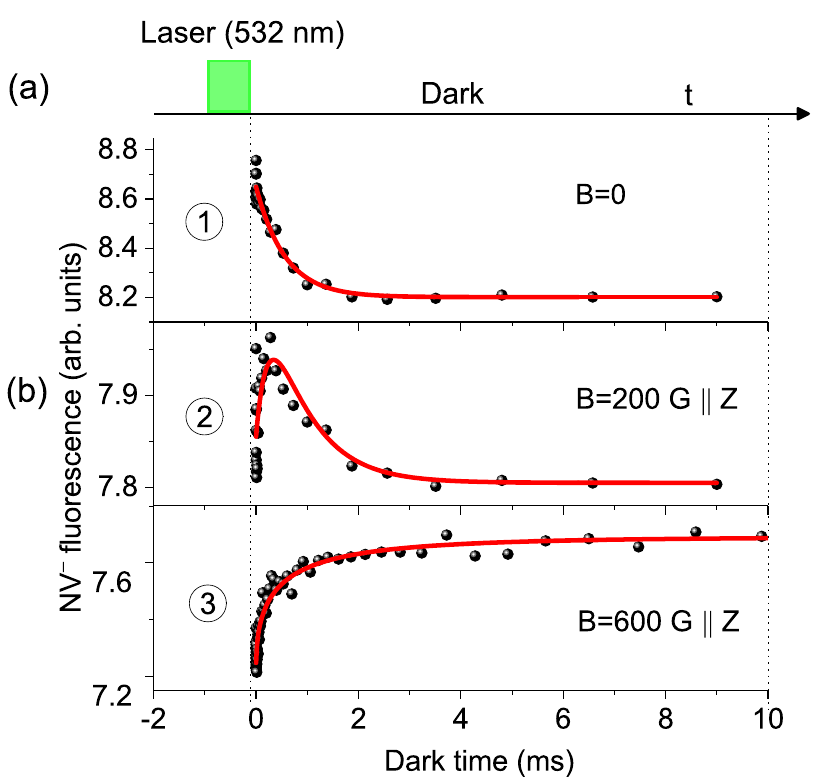}
  \caption{(a) Pulse sequence used to measure spin-charge dynamics in the dark following green excitation for 1 ms. (b) Fluorescence evolution with dark time for the three regimes marked in Fig. \ref{figure3a_3b}b. Regime-1 (B = 0): Charge dynamics is completely masked by spin relaxation in the dark. Regime-2 (B = 200 G): Reduction of spin polarization results in an almost equal spin-charge contribution to the fluorescence evolution in the dark. Regime-3 (B = 600 G): Spin polarization vanishes completely, revealing the hidden recharging process of NV centers in the dark.}
  \label{figure4a_4b}
\end{figure} 

Next, we demonstrate that an off-axis magnetic field could indeed quench the NV$^{-}$ spin polarization. We used the nanodiamond powder sample and employed a simple pulse sequence as shown in Fig. \ref{figure2a_2c}a (upper panel). The laser was off for the initial 5 ms, much longer than the typical T$_{1}$-relaxation time in these nanodiamonds, to allow the sample to relax in the dark \cite{Manson2006}. Then, we applied a weak 10-mW laser pulse at 532 nm, well below saturation, and monitored the fluorescence of the NV$^{-}$  centers (750-800 nm band). We show the time evolution of the fluorescence intensity in Fig. \ref{figure2a_2c}a (lower panel). At the beginning of the laser pulse, the fluorescence increases until it reaches a maximum and then starts to decay with time. We can interpret the evolution of the curve as follows. The laser pulse, due to the Gaussian nature of the excitation spot, polarizes the NDs with a characteristic time that depends on laser power and the intrinsic NV spin relaxation mechanisms. Also, ionization and recombination of the NV centers under green laser illumination are robust in the presence of high defect concentrations such as substitutional nitrogens \cite{Manson2005, JChen2017, Manson2018}.  These two competing processes result in the bi-exponential evolution of the fluorescence during illumination. This interplay between spin and charge dynamics has been thoroughly described in Ref. \cite{Giri2018}. As we discussed earlier, an off-axis magnetic can quench the NV$^{-}$ spin polarization. If the initial increase of the fluorescence is indeed due to spin polarization, then it should be affected by magnetic fields. Figure \ref{figure2a_2c}b shows that it is indeed the case. The rising component gets smaller in amplitude with increasing $\bold{B}$, and at 600 G the fluorescence evolution is dominated by a decaying component corresponding to increasing depletion of the NV$^{-}$ population by charge transfer. In order to substantiate this interpretation, we monitored the fluorescence of the NV$^{0}$ centers (550-600 nm band).  The time dependence of the NV$^{0}$  fluorescence intensity is symmetric to that of the NV$^{-}$  fluorescence intensity (Fig. \ref{figure2a_2c}c). This indicates that photoionization of NV$^{-}$ centers increases NV$^{0}$ populations which are reflected in their emission intensity \cite{Iakoubovskii2000, Kupriyanov2000, Gaebel2005, Manson2005}. The fluorescence signal is devoid of any contribution from spin polarization, and under this experimental condition, we can selectively explore the real charge state dynamics of the NV centers under green illumination.

We extracted the amplitudes and rates of the two competing processes of charge and spin dynamics during illumination, from the biexponential fits of curves as in Fig. \ref{figure2a_2c}a,b. The spin polarization rate and ionization rate as a function of the off-axis magnetic field are plotted in Fig. \ref{figure3a_3b}a. The spin polarization rate appears to increase as the magnetic field increases until spin-state mixing quenches the polarization completely (black spheres in Fig. \ref{figure3a_3b}a). The magnetic field can stimulate the transition from the NV$^{-}$  excited state to the metastable singlet state, resulting in an increase in the populations in the metastable state \cite{Manson2006, Capelli2017, Manson2018}, thereby increasing the rate at which the NV$^{-}$ centers are initialized into the $m_s=0$ state. However, this process can also induce the ionization of the NV$^{-}$ centers from the excited state as well as the metastable singlet state through a single-photon ionization process, a process typical of diamonds with high defect concentration \cite{Manson2006}. Thus, increasing the magnetic field strength leads to an increased rate of ionization (red hexagons in Fig. \ref{figure3a_3b}a).  However, the amplitude of the ionization process remains unaffected by the magnetic field (red hexagons in Fig. \ref{figure3a_3b}b), and the only noticeable effect of the magnetic field is a reduction of the amplitude of spin polarization (black spheres in Fig. \ref{figure3a_3b}b). Thus, we have three distinct regimes as marked in Fig. \ref{figure3a_3b}b. When no external magnetic field is applied, the spin polarization of the NV$^{-}$ centers dominates over the ionization process (Regime-1).  With an increase in the field strength, spin state mixing results in a decrease in the amplitude of spin polarization, and at about 200 G, the contributions of spin polarization, and ionization are almost equal (Regime-2). Any further increase in field strength results in a situation where the ionization process dominates over spin polarization, and at B= 600 G, the spin polarization is suppressed, revealing the full dynamics of the ionization process (Regime-3).

We further investigated the spin-charge dynamics in the dark in these three distinct regimes. We used a standard T$_{1}$-relaxation pulse sequence as shown in Fig. \ref{figure4a_4b}a. We initialized the system for 1 ms using a weak 532-nm laser and monitored the fluorescence evolution in the dark using a 1 $\mu$s read-out pulse at the same wavelength. For B=0, we observed a simple exponential type decay indicative of spin relaxation in the dark dominating over charge dynamics. The hidden charge dynamics was recently exploited to increase the sensitivity of magnetic noise detection \cite{Gorrini2018}, but it could also have a detrimental effect in the spin relaxation measurements \cite{Giri2018, Bluvstein2018}. At B= 200 G, charge dynamics emerge, as spin state mixing reduces the amplitude of spin polarization that could be achieved by the laser pulse. The result is a biexponential type decay curve similar to the ones previously reported \cite{Giri2018, Gorrini2018}. At B = 600 G, as the spin polarization is completely quenched, fluorescence evolution is dictated by charge dynamics in the dark. 

Thus, by quenching the spin polarization with a magnetic field, one can unravel the charge dynamics of the NV centers during illumination as well as in the dark. As a possible application to understand physical mechanisms governing charge state instability, we applied this technique to two systems.

\begin{figure}[!t]
\centering
  \includegraphics[width=1\columnwidth]{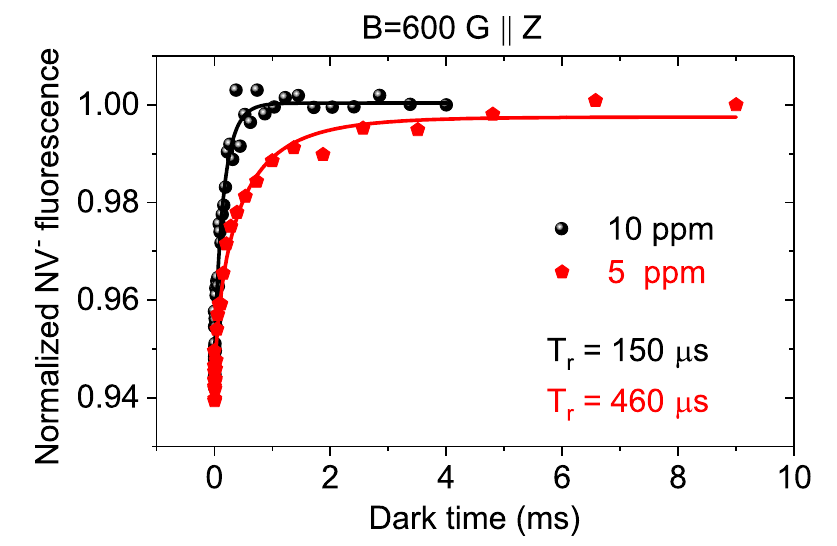}
  \caption{Charge state dynamics of NV centers in nanodiamonds in the dark, for two different NV density (5 ppm and 10 ppm). The fluorescence intensity is normalized to unity for t=t$_{max}$. The recharge-in-the-dark process becomes faster with increasing NV density.}
  \label{figure5}
\end{figure}

\textit{(i) Ensembles of NV centers in NDs.}
We compared the charge dynamics in the dark in two ND powder samples, measured at 600 G $\parallel$ Z using the T$_{1}$ sequence of Fig. \ref{figure4a_4b}a. The nominal diameter of the NDs is 100 nm. The primary difference between the two samples is the NV densities (one with 5 ppm and the other with 10 ppm of NV centers). The fluorescence intensity increased with dark time, reaching saturation (see Fig. \ref{figure5}). We can interpret this as follows.  During illumination NV$^{-}$s are ionized increasing the populations of NV$^{0}$s (Fig. \ref{figure2a_2c}c). In the dark, the NV$^{0}$ centers recharge due to electrons tunneling between proximate NV centers \cite{Choi2017}, or between NVs and nitrogen impurities \cite{Manson2018}, and establish a charge equilibrium condition. Thus NV$^{-}$ is the stable charge state in the NDs, despite large surface area in which the electron traps are known to favor the NV$^{0}$ charge state \cite{Rondin2010}. Increasing the NV density results in a shorter recharging time, indicating more charge state instability. This behavior could be due to an increase in the tunneling rates as the distance betwen the NV centers as well as between NVs and other defects decrease with increasing defect density \cite{Chou2018}.

\begin{figure}[!b]
\centering
  \includegraphics[width=1\columnwidth]{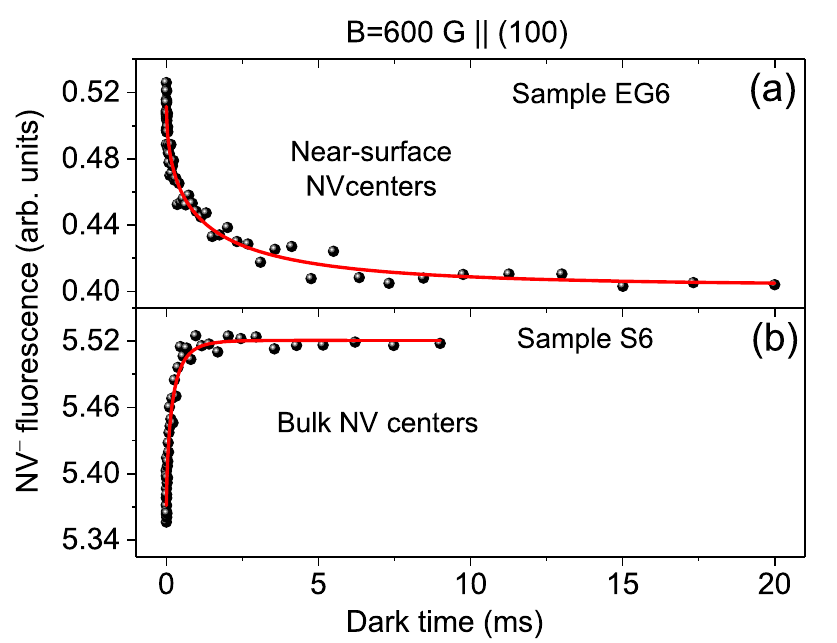}
  \caption{Charge state dynamics of (a) near-surface, and (b) bulk NV centers in the dark. The sign of fluorescence evolution in the dark is different for bulk and near-surface NV centers, indicating location dependent charge dynamics.}
  \label{figure6a_6b}
\end{figure} 

\textit{(ii) Bulk vs near-surface NV centers in bulk single crystal diamond.}
Recent work on single NV centers suggest different charge state dynamics for bulk and near-surface NV centers under green illumination \cite{BarGill2017}.  Here, we show that even in the case of ensembles of NVs, charge state dynamics of bulk- and the near-surface NV centers are different, both under green illumination as well as in equilibrium in the dark. Near-surface NV centers in a nitrogen-implanted  electronic grade CVD sample (EG6), and bulk NV centers from a non-electronic grade CVD sample (S6) were chosen for the study. We applied a magnetic field of 600 G along the [100] crystallographic direction to quench the spin polarization and used the T$_{1}$ sequence of Fig. \ref{figure4a_4b}a to investigate the charge dynamics in the dark. In the case of near-surface NV centers, in the steady-state under green excitation, photo-conversion results in an increased NV$^{-}$  population. In the dark, primarily the NV$^{-}$ centers ionize due to electron loss to nearby traps \cite{Bluvstein2018}, establishing an equilibrium condition, in which the NV$^{0}$ center is the stable charge state (see Fig. \ref{figure6a_6b}a). This result is a consequence of the fact that the diamond surface is known to favor the neutral charge state \cite{Santori2009, Fu2010, Hauf2011}. The dynamics are exactly opposite in the case of bulk NV centers and similar to the NDs, the equilibrium condition is the one in which NV$^{-}$ is the favored configuration (see Fig. \ref{figure6a_6b}b).

\section{Conclusions}
To summarize, we have shown that it is possible to completely decouple the effect of spin from charge dynamics measurements with an off-axis magnetic field, without substantially affecting their charge state stability. In 100 nm NDs, we demonstrate that despite their large surface area NV$^{-}$ is the favored charge state both under illumination as well as in the dark, and the charge state instability scales with the density of NV centers. This observation has direct implications in the use of NDs for sensing applications in biological systems, where highly fluorescent NDs are required \cite{Hsiao2015, Petrakova2015, Zhang2018, Robinson2018}. We also show that, in bulk single crystal diamonds, the charge stability depends on their location in the sample. In the bulk of the diamond, the NV centers are mostly stable in the negative charge state. However, the all-important near-surface NV centers are mostly stable in the undesirable neutral charge state. The novel non-perturbative technique to study charge state dynamics we demonstrated here could help better understand charge state dynamics due to tunneling to defect sites in high NV density nanodiamonds  in which the production process can generate high defect concentrations. It could also be useful to investigate the effect of surface states, local charge environments on the charge state stability. A better understanding of the processes could help to optimize the diamond surface to increase the NV$^{-}$ populations of the near-surface NV centers.




\end{document}